# Remarks on the lattice Green's Function for the anisotropic Face Centered Cubic Lattice

[††††]J. H. Asad[†] , R. S. Hijjawi [††], A. J. Sakaji[†††] and J. M. Khalifeh

[†] *Dep. of Physics, College of Sciences, Tabuk University.*
*P.O.Box.1144, Dhiba Road, Tabuk, Kingdom of Saudi Arabia.*
*e-mail:* jhasad1@yahoo.com.

*.Department of Physics, Mutah University, Jordan* [††]
*.e-Mail*: Hijjawi@mutah.edu.jo.

[†††] *Department of Physics, Ajman University, UAE.*
*e-Mail:* a_sakaji@yahoo.com.

*.Department of Physics, University of Jordan, Amman-11942, Jordan*[††††]
*e-mail:* jkalifa@ju.edu.jo

**Abstract**

An expression for the Green's function (GF) of anisotropic face centered cubic (IFCC) lattice is evaluated analytically and numerically for a single impurity problem. The density of states (DOS), phase shift and scattering cross section are expressed in terms of complete elliptic integrals of the first kind.

**Key words:** IFCC lattice, impurity, and Green's function.

# I-Introduction

The Lattice Green's Function (LGF) is defined as[1]

$$G(E) = \frac{\Omega}{(2\pi)^d} \int_{IBZ} \frac{F(k)}{E - E(k)} d\vec{k} \qquad (1.1)$$

where $E(k)$ is a dispersion relation, $F(k)$ is an appropriate function, $\Omega$ is the volume of the crystal in real space, $d$ is the dimension, and IBZ denotes that the integration is restricted to the first Brillouin zone[1,2].

The LGF is a basic function in the study of the solid state physics and condensed matter. It appears especially when impure solids are studied[3]. Green was the first physicist who established the basic concepts of Green's Function (GF) in the potential theory, and his work was focused on solving Laplace's and Poisson's equations with different boundary conditions. The use of GF method plays an important role in many-body problems[4], especially in problems of solid state physics where an enormous progress has been realized. In the mathematical problem of quantum theory which consists of solving linear operator equations with given boundary conditions, GF constitute the natural language to study boundary conditions.

Nowadays, GF is one of the most important concepts in many branches of physics, as many quantities in solid state physics can be expressed in terms of the LGF. In the following are some examples: statistical model of ferromagnetism such as Ising model[5], Heisenberg model [6], spherical model[7], random walk theory[8,9], diffusion[10], band structure[11], and analysis of infinite electric networks[12-18].

The LGF for several structure lattices has been widely studied during the second half of the last century. The LGF for the rectangular lattice has been investigated by Katsura and Inawashiro[19]; they used the Mellin-Barnes type integral. Recurrence relation, which gives the LGF along the diagonal direction from a couple of values of complete elliptic integrals of the first and second kinds for the rectangular and square lattices, has been derived by Morita[20].

The LGF for Simple Cubic (SC) lattice at the origin $G(0,0,0)$ has been investigated by many authors: Joyce[21] expressed $G(0,0,0)$ in terms of the complete elliptic integrals of the first kind, Horiguchi[22] expressed $G(1,0,0)$ as a sum of simple integrals of the complete elliptic integrals of the first kind and evaluated it numerically, Katsura et al[23-25]. investigated the LGF for many lattices. Recently, Glasser and Boersma[26] showed that $G(l,m,n)$ can be expressed rationally in terms of $G(0,0,0)$.

The LGF for the Face Centered Cubic (FCC) lattice was studied well

by Iwata[27], he expressed $G(0,0,0)$ in a compact form as a product of complete elliptic integrals of the first kind. The LGF at any lattice site $G(l,m,n)$ was studied by Mano[28]; $G(l,m,n)$ is expressed in terms of linear combinations of complete elliptic integrals of the first and second kind. In their paper Glasser and Boersma[26] expressed the LGF for FCC lattice rationally in terms of the known value of $G(0,0,0)$.

In a recent work we have evaluated analytically and numerically GF, density of states, phase shift, and scattering cross section for the following cases:

(i)  the Glasser cubic lattice[29],
(ii) the Body Centered Cubic lattice[30],
(iii) The general Glasser case[31],
(iv) The Face Center Cubic lattice[32].

In this paper we report on the single impurity lattice Green's function. The paper is organized as follows: Section II is devoted to the general definition of the diagonal lattice Green's function and its form inside and outside the band for the IFCC lattice in terms of complete elliptic integrals of the first kind. This section also contains the formulae for the density of states, phase shift and scattering cross section for a point defect case. In section III we present the results and discussion.

## II- The IFCC lattice Green's function

The diagonal Green's function for the IFCC lattice with nearest neighbor interaction is defined as[33-38]

$$G^o(L,L;E) = \frac{1}{\pi^3} \int_0^\pi \int_0^\pi \int_0^\pi \frac{dk_x dk_y dk_z}{E - a_1 Cos(k_x)Cos(k_y) - a_2 Cos(k_x)Cos(k_z) - a_3 Cos(k_y)Cos(k_z)}, \quad E \rangle 2a_1 + a_3 \quad (2.1)$$

Integrating the above equation and using the method of analytic continuation[36-38], the diagonal Green's function outside the band has the form

$$G^0(L,L;E) = \frac{4}{\pi^2(E+a_3)} K(k_+) K(k_-), \quad E > 2a_1 + a_3 \quad (2.2)$$

where

$$k_{\pm}^2 = \frac{1}{2}\left(1 \mp 4\frac{\sqrt{\frac{a_3}{a_1^2}E(1+\frac{a_3}{a_1^2}E)}}{(E+a_3)^2} - \frac{\sqrt{(E+a_3)^2 - 4\frac{a_3}{a_1^2}E}\sqrt{(E+a_3)^2 - 4(1+\frac{a_3}{a_1^2}E)}}{(E+a_3)^2}\right), \qquad (2.3)$$

Green's function outside and inside the band can be written as (all mathematical Manipulations are given in appendix A).

$$G^o(L,L;E) = \begin{cases} \dfrac{4}{\pi^2(E+a_3)} K(k_+)K(k_-), & E > 2a_1+a_3 \\ \dfrac{2}{\pi^2(E+a_3)}[(X_+^2+1)(X_-^2+1)]^{-\frac{1}{4}}(K(v_+)K(u_-)+K(v_-)K(u_+)+i[K(v_+)K(u_+)-K(v_-)K(u_-)]), & -(2a_1-a_3)<E<0 \end{cases}, (2.4)$$

where

$$X_{\mp} = \frac{\sqrt{-E}}{(E+a_3)^2}\left(-\sqrt{\frac{[(E+a_3)^2 - 4\frac{a_3}{a_1^2}E][4(1+\frac{a_3}{a_1^2}E)-(E+a_3)^2]}{-E}} \mp 4\sqrt{\frac{a_3}{a_1^2}(1+\frac{a_3}{a_1^2}E)}\right), \qquad (2.5)$$

and

$$v_\pm^2 = \frac{1}{2}\left(1 \pm \sqrt{\frac{X_-^2}{X_-^2+1}}\right) \qquad (2.6)$$

$$u_\pm^2 = \frac{1}{2}\left(1 \pm \sqrt{\frac{X_+^2}{X_+^2+1}}\right) \qquad (2.7)$$

Now the density of states is defined as:

$$DOS(E) = \frac{1}{\pi}\operatorname{Im} G \qquad (2.8a)$$

$$DOS^o(E) = \frac{2}{\pi^3(E+a_3)}[(X_-^2+1)(X_+^2+1)]^{-\frac{1}{4}}[K(v_+)K(u_+)-K(v_-)K(u_-)], \quad -(2a_1-a_3)<E<0 \qquad (2.8b)$$

where $K(v_\pm)$ and $K(u_\pm)$ are the complete elliptic integrals of the first kind.

Consider the case of a tight-binding Hamiltonian whose perfect periodicity is destroyed due to the presence of the point defect at the L site. This situation can be thought of physically as arising by substituting the host atom at the L-site by a foreign atom[1,39] having a level lying $\varepsilon'$ higher than the common level of the host atoms (L). Normally, this atom is close to the host in the same series of the periodic table.

Thus, our diagonal Green's function of the IFCC lattice for the single impurity case can be written as

$$G(L,L,E) = \begin{cases} \dfrac{4K(k_+)K(k_-)}{\pi^2(E+a_3) - 4\varepsilon' K(k_+)K(k_-)} \quad ; \quad E > 2a_1 + a_3 & (2.9) \\[2ex] \dfrac{\dfrac{\pi}{2}(E+a_3)[(X_+^2+1)(X_-^2+1)]^{\frac{1}{4}}[K(v_+)K(u_-) + K(v_-)K(u_+) + i(K(v_+)K(u_+) - K(v_-)K(u_-))] - \varepsilon'[K^2(v_+) + K^2(v_-)][K^2(u_+) + K^2(u_-)]}{[\dfrac{\pi^2}{2}(E+a_3)((X_+^2+1)(X_-^2+1))^{\frac{1}{4}} - \varepsilon'(K(v_+)K(u_-) + K(v_-)K(u_+))]^2 + \varepsilon'^2[K(v_+)K(u_+) - K(v_-)K(u_-)]^2}, \; -(2a_1 - a_3) < E < 0 \end{cases}$$

and the corresponding density of states can be written as:

$$DOS(E) = \frac{\dfrac{\pi}{2}(E+a_3)[(X_+^2+1)(X_-^2+1)]^{\frac{1}{4}}(K(v_+)K(u_-) - K(v_-)K(u_+)}{[\dfrac{\pi^2}{2}(E+a_3)((X_+^2+1)(X_-^2+1))^{\frac{1}{4}} - \varepsilon'(K(v_+)K(u_-) + K(v_-)K(u_+))]^2 + \varepsilon'^2[K(v_+)K(u_+) - K(v_-)K(u_-)]^2}, \; -(2a_1 - a_3) < E < 0 \quad (2.10)$$

The S-wave phase shift, $\delta_0$, is defined as[39]:

$$\tan\delta_o = \frac{\operatorname{Im} G^0(E)}{\dfrac{1}{\varepsilon'} - \operatorname{Re} G^0(E)}, \quad (2.11)$$

Here, $\operatorname{Re} G^0(E)$ and $\operatorname{Im} G^0(E)$ refer to the real and imaginary parts of the Green's function inside the band respectively. After some mathematical manipulations, we obtain:

$$\tan\delta_o = \frac{K(v_+)K(u_+) - K(v_-)K(u_-)}{\dfrac{\pi^2(E+a_3)[(X_+^2+1)(X_-^2+1)]^{\frac{1}{4}}}{2\varepsilon'} - (K(v_+)K(u_-) + K(v_-)K(u_+))}, \quad (2.12)$$

The cross-section, $\sigma$, is defined as [39]:

$$\sigma = \frac{4\pi}{P^2} \frac{[\operatorname{Im} G^0(E)]^2}{\left[\operatorname{Re} G^0(E) - \dfrac{1}{\varepsilon'}\right]^2 + [\operatorname{Im} G^0(E)]^2}, \quad (2.13)$$

Here, P refers to the electron momentum.

Therefore, the cross-section becomes

$$\sigma = \frac{4\pi}{P^2} \frac{[K(v_+)K(u_+) - K(v_-)K(u_-)]^2}{\left[K(v_+)K(u_-) + K(v_-)K(u_+) - \dfrac{\pi^2(E+a_3)[(X_+^2+1)(X_-^2+1)]^{\frac{1}{4}}}{2\varepsilon'}\right]^2 + [K(v_+)K(u_+) - K(v_-)K(u_-)]^2}.$$

Integrating the above equation and using the method of analytic continuation [36-38], the diagonal Green's function outside the band has the form

Special cases:

(i)     When $a_1 = a_1 = a_3 = 1$ we find face centered cubic (FCC) lattice.

(ii)     When $a_1 = 1$ and $a_3 = 0$

The diagonal Green's function outside the band has the form

$$G^0(L,L;E) = \frac{4}{\pi^2 E} K^2(k), \qquad |E| > 2 \qquad (2.2')$$

where

$$k^2 = \frac{1}{2}\left(1 - \sqrt{1 - \frac{4}{E^2}}\right), \qquad (2.3')$$

Green's function outside and inside the band can be written as (all mathematical Manipulations are given in appendix A).

$$G^o(L,L,;E) = \begin{cases} \dfrac{4}{\pi^2 E} K^2(k), & |E| > 2 \\ \dfrac{1}{\pi^2}(2K(u_+)K(u_-) + i[K^2(u_+) - K^2(u_-)]), & -2 < E < 0 \end{cases}, \quad (2.4')$$

where

$$X_+ = X_- = \left(-\sqrt{\dfrac{4-E^2}{E^2}}\right), \quad (2.5')$$

and

$$u_\pm^2 = v_\pm^2 = \dfrac{1}{2}\left(1 \pm \sqrt{1 - \dfrac{E^2}{4}}\right) \quad (2.6',7')$$

Therefore, the density of states is

$$\text{DOS}^o(E) = \dfrac{1}{\pi^3}[K^2(u_+) - K^2(u_-)], \quad -2 < E < 0 \quad (2.8')$$

where $K(v_\pm)$ and $K(u_\pm)$ are the complete elliptic integrals of the first kind.

Thus, our diagonal Green's function of the IFCC lattice for the single impurity case can be written as

$$G(L,L,E) = \begin{cases} \dfrac{4K^2(k)}{\dfrac{\pi^2 E}{4} - \varepsilon' K^2(k)}; & E > 2 \quad (2.9') \\ \dfrac{\pi^2[2K(u_+)K(u_-) + i(K^2(u_+) - K^2(u_-))] - \varepsilon'[K^2(u_+) + K^2(u_-)]^2}{[\pi^2 - 2\varepsilon' K(u_+)K(u_-)]^2 + \varepsilon'^2[K^2(u_+) - K^2(u_-)]^2}, & -2 < E < 0 \end{cases}$$

and the corresponding density of states can be written as:

$$DOS(E) = \frac{\pi (K^2(u_+) - K^2(u_-))}{[\pi^2 - 2\varepsilon' K(u_+)K(u_-)]^2 + \varepsilon'^2 [K^2(u_+) - K^2(u_-)]^2}, \quad -2 < E < 0 \tag{2.10'}$$

The S-wave phase shift, $\delta_0$, is:

$$\tan\delta_o = \frac{K^2(u_+) - K^2(u_-)}{\frac{\pi^2}{\varepsilon'} - 2K(u_+)K(u_-)}, \tag{2.12'}$$

The cross-section, $\sigma$, is:

$$\sigma = \frac{4\pi}{P^2} \frac{[K^2(u_+) - K^2(u_-)]^2}{\left[2K(u_+)K(u_-) - \frac{\pi^2}{\varepsilon'}\right]^2 + [K^2(u_+) - K^2(u_-)]^2}. \tag{2.14'}$$

**III- Results and Discussion**

Our results for the body centered cubic lattice are shown in Figures (1-9). Figures (1,2) show real and imaginary parts of Green's Function for the pure lattice. The figures show logarithmic behavior. Fig [3] shows the density of states for the pure lattice. The density of states has the same behavior as above apart from a constant. The figure shows that the function is symmetric (even function).

Figure 4 shows the density of states for the body centered cubic lattice with single impurity for different potential strengths $\varepsilon'$ (-0.6,-0.3, 0.0, 0.3, and 0.6). For $\varepsilon' = 0.0$ it falls off exponentially. The peak value varies with the potential strengths and reaches its maximum at $\varepsilon' = 0.3$, also the divergence of the density of states removed by adding such impurities. Figure 5 shows the density of states for the body centered cubic lattice (DOS) in three-dimensions with one axis representing potential strengths $\varepsilon'$ varying between −1 and 1 (arbitrary units) whereas the second axis is energy scale varying between −1 and 1 as indicated in the formalism.

The phase shift, $\delta_0$, is defined as the shift in the phase of the wave function due to the presence of the impurity potential. Figure 6 displays, $\delta_0$, for the body centered cubic lattice with single impurity for different potential strengths $\varepsilon'$ (-0.6,-0.3, 0.0, 0.3, and 0.6). For $\varepsilon' = 0.0$, $\delta_0$ vanishes as potential is turned off (perfect lattice). The phase shift is always negative for all negative potential strengths $\varepsilon'$. In range between $\varepsilon'=0.00$ and $\varepsilon'=0.3$, $\delta_0$ is positive. In the range $\varepsilon'$ varies between 0.3 and 1.0 we have discontinuity as shown in Fig.6, $\delta_0$, displays into two regions around the discontinuity point, right hand region is negative and it increases if $\varepsilon'$ increases, the left hand region is positive and it decreases if $\varepsilon'$ increases (discontinuity point moves to the left). Figure 7 shows the phase shift, $\delta_0$, in three dimensions for the body centered cubic lattice with single impurity for different potential strengths $\varepsilon'$ varying between -1 and 1 (arbitrary units).

The cross section, $\sigma$, is defined as the area an impurity atom presents to the incident electron. Figure 8 shows the cross section for single substitutional impurity with different potential strength $\varepsilon'$, the peak value varies with the potential strength and reaches its maximum for all values $\varepsilon' > 0.3$, in the range $\varepsilon'$ varies between 0.0 and 0.3 the peak value increases if $\varepsilon'$ increases, in range between $\varepsilon'$ 0.0 and –1.0 the peak value increases if $\varepsilon'$ decreases. The values are all positive since $\sigma$ can be viewed as a sort of probability. It is related to some physical quantities such as the conductivity in metals. Figure 9 shows the cross section, $\sigma$, in three dimensions for the body centered cubic lattice with single impurity for different potential strengths $\varepsilon'$ varying between -1 and 1(arbitrary units).

# Appendix A

## Derivation of Green's function for the face centered cubic lattice inside the band

In this Appendix we derive an expression for Green's function inside the band in terms of complete elliptic integral of the first kind.
Green's function for the face centered cubic lattice outside the band is given by[10-16]:

$$G^0(L,L;E) = \frac{4}{\pi^2(E+a_3)} K(k_+)K(k_-), \qquad E > 2a_1 + a_3 \qquad (A.1)$$

where

$$k_\pm^2 = \frac{1}{2}\left(1 \mp 4\frac{\sqrt{\frac{a_3}{a_1^2}E(1+\frac{a_3}{a_1^2}E)}}{(E+a_3)^2} - \frac{\sqrt{(E+a_3)^2 - 4\frac{a_3}{a_1^2}E}\sqrt{(E+a_3)^2 - 4(1+\frac{a_3}{a_1^2}E)}}{(E+a_3)^2}\right), \qquad (A.2)$$

Or in the range E enclosed between $-(2a_1-a_3)$ and 0

$$k_\pm^2 = \frac{1}{2}(1+Z_\mp), \qquad -(2a_1-a_3) < E < 0 \qquad (A.3)$$

$$Z_\mp = iX_\mp \qquad (A.4)$$

where

$$X_\mp = \frac{\sqrt{-E}}{(E+a_3)^2}\left(-\sqrt{\frac{[(E+a_3)^2 - 4\frac{a_3}{a_1^2}E][4(1+\frac{a_3}{a_1^2}E) - (E+a_3)^2]}{-E}} \mp 4\sqrt{\frac{a_3}{a_1^2}(1+\frac{a_3}{a_1^2}E)}\right), \qquad (A.5)$$

The complete elliptic integral of the first kind is expressed as

$$K(k) = \frac{\pi}{2}{}_2F_1(\frac{1}{2},\frac{1}{2},1,k^2) \qquad (A.6)$$

where

$_2F_1(\frac{1}{2},\frac{1}{2},1,k^2)$ is the Gauss hypergeometric function

Substituting (A.6) in (A.1) we have

$$G^0(E) = \frac{{}_2F_1(\frac{1}{2},\frac{1}{2};1;k_+^2)\,{}_2F_1(\frac{1}{2},\frac{1}{2};1;k_-^2)}{E + a_3} \quad (A.7)$$

:Using the following transformations[19]

$${}_2F_1(\frac{1}{2},\frac{1}{2};1;\frac{1+Z_\mp}{2}) = \frac{\Gamma(\frac{1}{2})}{(\Gamma(\frac{3}{4}))^2}\,{}_2F_1(\frac{1}{4},\frac{1}{4};\frac{1}{2};Z_\mp^2) + 2Z_\mp \frac{\Gamma(\frac{1}{2})}{(\Gamma(\frac{1}{4}))^2}\,{}_2F_1(\frac{3}{4},\frac{3}{4};\frac{3}{2};Z_\mp^2), \quad (A.8)$$

With

$${}_2F_1(a,b;c;Z_\mp^2) = (1-Z_\mp^2)^{-a}\,{}_2F_1(a,c-b;c;\frac{Z_\mp^2}{Z_\mp^2-1}) \quad (A.9)$$

$$\frac{2\Gamma(\frac{1}{2})}{(\Gamma(\frac{3}{4}))^2}\,{}_2F_1(\frac{1}{4},\frac{1}{4};\frac{1}{2};\frac{Z_\mp^2}{Z_\mp^2-1}) = {}_2F_1(\frac{1}{2},\frac{1}{2};1;\frac{1}{2}(1+\sqrt{\frac{Z_\mp^2}{Z_\mp^2-1}})) + {}_2F_1(\frac{1}{2},\frac{1}{2};1;\frac{1}{2}(1-\sqrt{\frac{Z_\mp^2}{Z_\mp^2-1}})), \quad (A.10)$$

$$\frac{2\Gamma(-\frac{1}{2})}{(\Gamma(\frac{1}{4}))^2}\sqrt{\frac{Z_\mp^2}{Z_\mp^2-1}}\,_2F_1(\frac{3}{4},\frac{3}{4};\frac{3}{2};\frac{Z_\mp^2}{Z_\mp^2-1})=\,_2F_1(\frac{1}{2},\frac{1}{2};1;\frac{1}{2}(1-\sqrt{\frac{Z_\mp^2}{Z_\mp^2-1}}))\ -\ _2F_1(\frac{1}{2},\frac{1}{2};1;\frac{1}{2}(1+\sqrt{\frac{Z_\mp^2}{Z_\mp^2-1}})), \quad (A.11)$$

Substituting (A.8), (A.9), (A.10) and (A.11) in (A.7) then we

$$G^0(L,L,E) = \frac{2}{\pi^2(E+a_3)}[(X_+^2+1)(X_-^2+1)]^{\frac{-1}{4}}(K(v_+)K(u_-)+K(v_-)K(u_+)+i(K(v_+)K(u_+)-K(v_-)K(u_-))), \quad (A.12)$$

where

$$v_\pm = \frac{1}{2}(1\pm\sqrt{\frac{X_-^2}{X_-^2+1}}) \qquad (A.13)$$

$$u_\pm = \frac{1}{2}(1\pm\sqrt{\frac{X_+^2}{X_+^2+1}}) \qquad (A.14)$$

If we have a single impurity then Green's function is defined as[1]:

$$G(L,L,E) = \frac{G^0(L,L,E)}{1-\varepsilon'G^0(L,L,E)} \qquad (A.15)$$

.After some mathematical manipulation Eq. (A.15) becomes

$$G(L,L,E) = \frac{\frac{\pi^2}{2}(E+a_3)[(X_+^2+1)(X_-^2+1)]^{\frac{1}{4}}[K(v_+)K(u_-)+K(v_-)K(u_+)+i(K(v_+)K(u_+)-K(v_-)K(u_-))] - \varepsilon'[K^2(v_+)+K^2(v_-)][K^2(u_+)+K^2(u_-)]}{(\frac{\pi^2}{2}(E+a_3)((X_+^2+1)(X_-^2+1))^{\frac{1}{4}} - \varepsilon'(K(v_+)K(u_-)+K(v_-)K(u_+)))^2 + \varepsilon'^2(K(v_+)K(u_+)-K(v_-)K(u_-))^2}. \quad (A.16)$$

Thus, the S-phase shift, and scattering cross section can be evaluated in terms of complete elliptic integrals of the first kind as shown in the text.

# Figure Captions

Fig. 1: Real part Green's Function for the perfect FCC lattice.
Fig. 2: Imaginary part Green's Function for the perfect FCC lattice.
Fig. 3: The density of states for the perfect FCC lattice.
Fig. 4: The density of states (DOS) for the FCC lattice with single impurity for different potential strengths $\varepsilon'$ (-0.6,-0.3,0.0,0.3, and 0.6).
Fig.5: Three-dimensional density of states (DOS) for the FCC lattice with single impurity for different potential strengths $\varepsilon'$ varying between -1 and 1(arbitrary units ).
Fig. 6: The phase shift, $\delta_0$, for the FCC lattice with single impurity for different potential strengths $\varepsilon'$ (-0.6,-0.3,0.0,0.3, and 0.6).
Fig. 7: The phase shift, $\delta_0$, in three dimensions for the FCC lattice with single impurity for different potential strengths $\varepsilon'$ varying between -1 and 1(arbitrary units ).
Fig. 8: The cross section, $\sigma$, for the FCC lattice with single impurity for different potential strengths $\varepsilon'$ (-0.6,-0.3,0.0,0.3, and 0.6).
Fig. 9: The cross section, $\sigma$, in three dimensions for the FCC lattice with single impurity for different potential strengths $\varepsilon'$ varying between -1 and 1(arbitrary units).

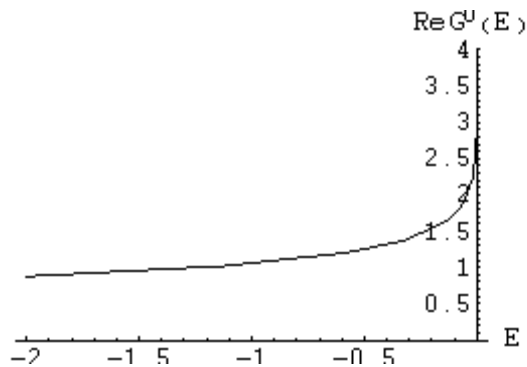

Fig.1

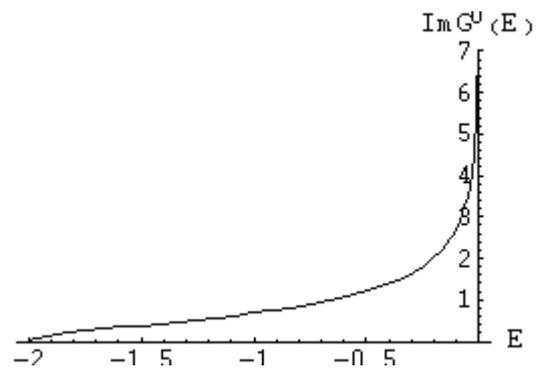

Fig. 2

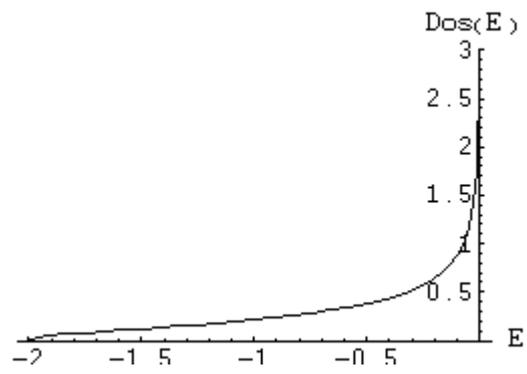

Fig. 3

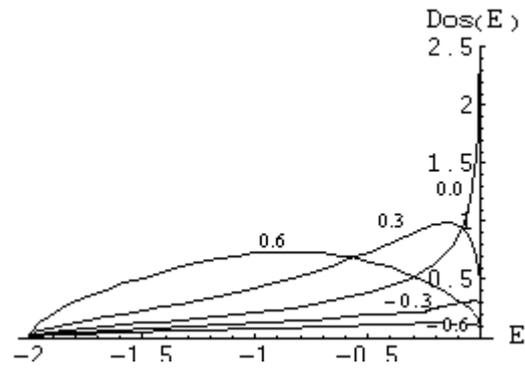

Fig. 4

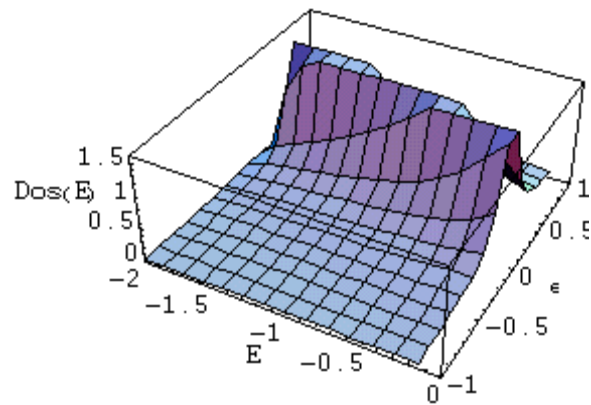

Fig. 5

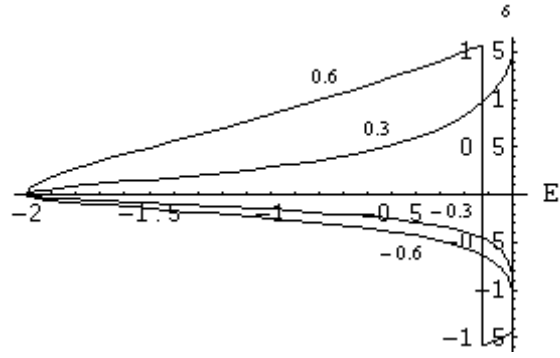

Fig.6

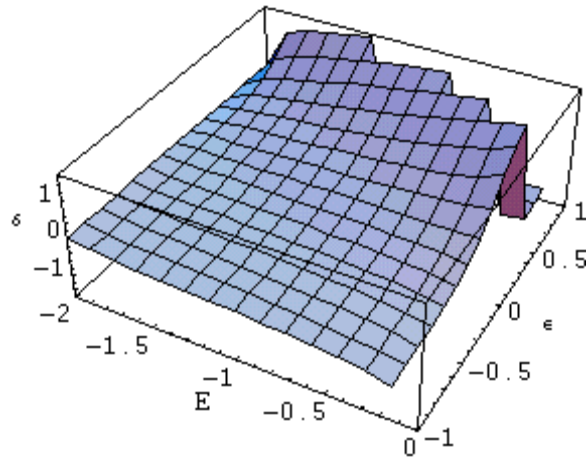

Fig. 7

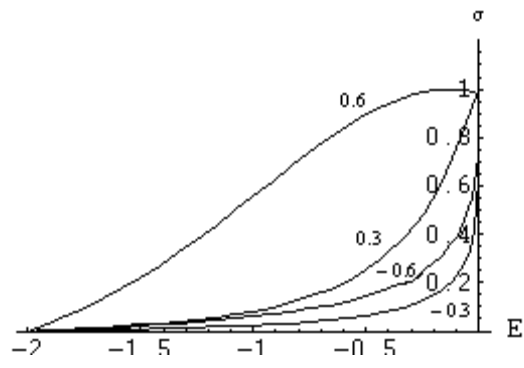

Fig. 8

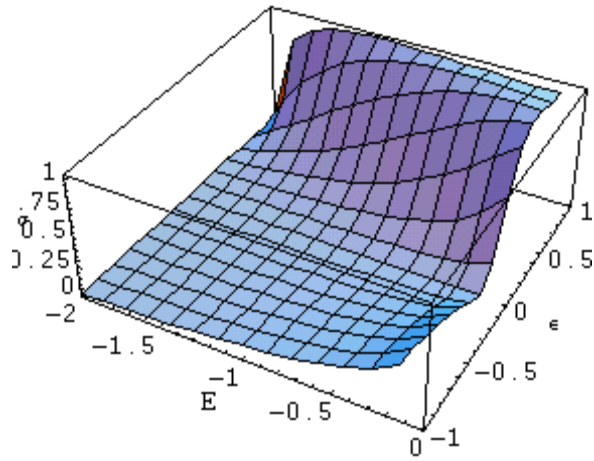

Fig. 9